# Ambient-pressure 151-K superconductivity in $HgBa_2Ca_2Cu_3O_{8+\delta}$ *via* pressure quench

Liangzi Deng[1,*], Thacien Habamahoro[1], Artin Safezoddeh[1], Bishnu Karki[1], Sudaice Kazibwe[1], Daniel J. Schulze[1], Zheng Wu[1], Matthew Julian[2], Rohit P. Prasankumar[2], Hua Zhou[3], Jesse S. Smith[3], Pavan R. Hosur[1], Ching-Wu Chu[1,*]

[1]Department of Physics and Texas Center for Superconductivity at the University of Houston (TcSUH), Houston, Texas 77204, USA

[2]Enterprise Science Fund, Intellectual Ventures, Bellevue, Washington 98005, USA

[3]X-ray Science Division, Argonne National Laboratory, Lemont, Illinois 60439, USA

*Liangzi Deng, Ching-Wu Chu

**Email:** cwchu@uh.edu, ldeng2@central.uh.edu

**Author Contributions:** L.Z.D. and C.W.C. designed research; L.Z.D., T.H., A.S., S.K., D.J.S., H.Z., and J.S.S. performed research; B.K. and P.H. contributed new analytic tools; L.Z.D., Z.W., M.J., R.P.P., and C.W.C. analyzed data; and L.Z.D. and C.W.C. wrote the paper.

**Competing Interest Statement:** The authors declare no competing interests.

**Classification:** Physical Sciences: Physics

**Abstract**

Superconductivity has been a vigorously researched topic since its discovery in 1911. Raising the superconducting transition temperature ($T_c$) has been the main driving force behind such long-sustained efforts due to its potential for impacting humanity and the fundamental knowledge gained from understanding this macroscopic coherent quantum state at high temperatures. The successful development of high-$T_c$ superconductivity will make possible extraordinarily efficient generation, delivery, and utilization of energy, and could also enable the development of controlled fusion while impacting other burgeoning fields like quantum computation and quantum electronics. However, progress has been hindered by a longstanding plateau in the record ambient-pressure $T_c$, unchanged since 1993. Subsequent significant advancements in $T_c$ have been achieved only under high pressures, preventing the realization of superconductivity's full potential. To directly address this challenge, we developed a pressure-quench protocol (PQP) to stabilize pressure-induced/-enhanced superconducting states at ambient pressure. Here we achieve a record ambient-pressure $T_c$ of 151 K in the cuprate $HgBa_2Ca_2Cu_3O_{8+\delta}$ *via* PQP. The experimental results are further supported by synchrotron X-ray diffraction measurements and phonon and electronic structure calculations. This breakthrough opens new avenues for stabilizing and exploring ambient-pressure high-$T_c$ superconducting states and other quantum states that have been previously only accessible under pressure, paving the way for deeper understanding and practical applications of high-$T_c$ superconductivity and beyond.

**Significance Statement**

The pressure-quench protocol (PQP) demonstrated here establishes a paradigm for stabilizing at ambient pressure the high-pressure–induced/–enhanced metastable phases that host elevated superconducting transition temperatures, an effective way to achieve record ambient-pressure high-temperature superconductivity. Its applicability extends well beyond superconductivity: PQP provides a powerful route to preserve quantum states that exist only under extreme conditions initially, making them accessible to advanced experimental probes—including a wide range of microscopic spectroscopies—under ambient environments. This capability opens pathways to investigate previously inaccessible physical phenomena and bridges the gap between fundamental discoveries and practical technologies. Moreover, PQP represents a new non-equilibrium strategy for uncovering novel states, including those absent not only at ambient pressure but even within the high-pressure regime originally.

## Main Text

## Introduction

Great progress has been made in raising $T_c$ (1-15), especially over the last four decades. By exploring different materials, $T_c$ rose from 4.2 K in the relatively inaccessible liquid-helium regime for elemental Hg in 1911 (1), to 23.2 K in the more accessible liquid-hydrogen regime for thin films of the A15 compound $Nb_3Ge$ in 1974 (16), and to 93 K in the industrial liquid-nitrogen regime for the perovskite-like cuprate $YBa_2Cu_3O_7$ in 1987 (17), ultimately reaching 133 K for the cuprate $HgBa_2Ca_2Cu_3O_{8+\delta}$ (Hg1223) in 1993 (8). The high $T_c$ in this compound reflects enormous pairing gaps in the range of 45-70 meV as seen in scanning tunneling microscopy (STM) measurements reported recently (18). This has remained the record-high $T_c$ at ambient pressure for over three decades. Under pressure, the $T_c$ of Hg1223 was further increased to 164 K at 31 GPa without any structural phase transition in 1994 (11), a temperature achievable in the cargo bay of the International Space Station facing away from the sun (19). The highest $T_c$ to date, up to 260 K, was reported in $LaH_{10}$ under a pressure of 190 GPa in 2019 (14). This rapid rise in $T_c$ in recent years is indeed impressive. It would have been even more so if the high pressure required to achieve these high-$T_c$ metastable states could have been lowered to ambient pressure to realize the full potential of superconductivity.

The unusually high $T_c$ and very large positive pressure effect on $T_c$ in the optimally doped n = 3 member of the cuprate homologue series $HgBa_2Ca_{n-1}Cu_nO_{2n+2+\delta}$ (HBCCO), *i.e.* Hg1223, can be qualitatively understood in terms of the high electron density of states of a possible van Hove singularity associated with the two-dimensional $CuO_2$ planes in HBCCO (20) and the possible suppression, by pressure and/or doping, of the many highly degenerated ground states unfavorable to high $T_c$ that are predicted to exist in strongly correlated electron systems such as cuprates. We conjecture that such an electronic van Hove singularity and/or other anomalies in the electronic energy spectrum, *e.g.* a Fermi surface topology change with a 2½-order transition (21-23), may facilitate the trapping of the high-pressure-induced metastable superconducting phase at ambient pressure, in the absence of a structural transition under pressure. As will be shown below, this is in qualitative agreement with our preliminary density functional theory band structure calculations up to 50 GPa.

Among the chemically stable cuprate homologue series HBCCO, Hg1223 displays the highest $T_c$ of 133 K in its pristine form and 164 K under 31 GPa. We therefore decided to achieve a new record ambient-pressure $T_c$ by stabilizing the high-pressure-induced metastable high-$T_c$ phase in Hg1223 at ambient pressure with a $T_c$ above 133 K by PQP. Recently, we have successfully demonstrated that, when anomalies exist in the phonon and/or electronic energy spectra, the pressure-induced/-enhanced metastable superconducting phases in *e.g.* Sb (24), FeSe (25-27), Cu-doped FeSe (25-27), and $Bi_{0.5}Se_{1.5}Te_3$ (28), can be retained at ambient pressure *via* the PQP we developed. Defects might also play an important role in the stabilization of metastable states during PQP, such as has been demonstrated in ion implantation in semiconductor device processing (29). The procedure for PQP is illustrated in Fig. 1A. Three critical parameters in this process are the quenching pressure ($P_Q$), the pressure from which the system is quenched (*i.e.* rapidly released) to ambient pressure; the quenching temperature ($T_Q$), the temperature at which pressure quenching takes place; and the speed of pressure quenching to ambient ($dP_Q/dt$, *i.e.* $v_Q$). Our PQP consists of three primary stages: (I) the creation and identification under pressure of the target phase to be retained ; (II) the rapid removal of pressure, followed by the immediate confirmation of both the retention of the targeted phase and its stability inside the diamond anvil cell (DAC); and (III) the retrieval of the pressure-quenched (PQed) sample from the DAC with minimum disturbance and its subsequent characterization outside the DAC *via* systematic studies at ambient pressure within the stability limit of the PQed phase.

Here, we report a new record-high ambient-pressure $T_c$ up to 151 K achieved in Hg1223 by employing the PQP at $P_Q$ = 10–30 GPa and $T_Q$ = 4.2 K and discuss possible origins for the retention of this high-pressure-induced high $T_c$ metastable phase at ambient pressure. A lower ambient-pressure $T_c$ of 139 K was also obtained by PQP at $P_Q$ = 26.3 GPa but at a higher $T_Q$ = 77 K. Our preliminary synchrotron X-ray diffraction (XRD) measurements conducted at the Advanced Photon Source (APS) at Argonne National Laboratory (ANL) on a sample after PQ at 30.9 GPa and 4.2 K with minimum disturbance do not indicate a structural change but clearly show a broadening of the diffraction lines at ambient pressure. We therefore currently attribute the retention of the PQed phase to strain and/or structural defects, with details yet to be resolved. This is consistent with our stability investigation, in which the PQed high-$T_c$ phase is annealed away through thermal cycling to higher temperature. The success of this experiment demonstrates the potential of the PQ technique for retaining pressure-induced metastable phases in solids with other useful and significant properties for science and technology, besides superconductivity, by taking advantage of anomalies in the phonon and/or electron energy spectra of solids at ambient pressure.

## Results and Discussion

### Determination of Target Phases in Hg1223 for PQP

To establish a new record-high $T_c$ at ambient pressure *via* PQP, we have chosen the chemically stable Hg1223 since it has the current record-high $T_c$ of 133 K at ambient pressure and, exhibiting a large positive pressure effect on its $T_c$, 164 K under pressure (11). The basic idea in this study is to bring the $T_c$ of Hg1223 to a higher value, namely the target $T_c$, by the application of pressure and then retain it at ambient pressure upon the rapid withdrawal of pressure, according to the PQP, taking advantage of the anomalous electronic energy spectrum of Hg1223 to be discussed later. Five samples (S1–S5) with size ~ 100 – 130 μm were cut from three self-flux-grown Hg1223 single crystals (S1, S2, and S4 from crystal #1, S3 from crystal #2, and S5 from crystal #3) for investigation. Detailed synthesis and annealing conditions can be found in "Methods – Sample Preparation". The X-ray diffraction (XRD) and temperature-dependent magnetic susceptibility $\chi(T)$ data for a representative sample prior to pressurization are shown in the left and right insets, respectively, in Fig. 1B. The high quality of the samples investigated is evident from the narrow and sharp Bragg peaks as well as the superconducting transition at 133 K. To implement the PQP, we first determined the pressure-dependent onset $T_c$, defined as the temperature where dR/dT rises rapidly on cooling (displayed *e.g.* in the inset to Fig. 2A), of Hg1223. As shown in Fig. 1B, the $T_c$ of our samples rises rapidly with pressure from 133 K to > 150 K above 10 GPa, in general agreement with previous reports (9-12). Slight variations in $T_c$ among the different studies exist, perhaps due to possible differences in doping of the samples examined, the pressure-transmitting medium used, or the definition of $T_c$ adopted (*i.e.* zero-resistance, mid-point, or onset $T_c$).

Since the $T_c$s of the metastable phases of Hg1223 under high pressures reach > 150 K, *i.e.* higher than that at ambient pressure (Fig. 1B and its right inset), it is natural for us to subject these phases with $T_c$s higher than 150 K (*i.e.* the target $T_c$s) for PQP, as indicated by the arrows in Fig. 1B, to reach record-breaking $T_c$s at ambient pressure. The results of six representative experiments at different $P_Q$s and $T_Q$s prior to and after PQP are presented and discussed below.

### Pressure Quenching Hg1223

Typical temperature-dependent resistance R(T) curves for Hg1223 samples S1–S4 under different conditions before and after PQP are displayed in Fig. 2, with each $T_c$ cited hereafter being the onset critical transition temperature as defined in the previous section and determined by the temperature-dependent dR/dT results for the R(T) curves in Fig. 2, as shown in the inset to Fig. 2A and in Fig. S1. Overall, $T_c$s ranging from 139 K to 151 K were achieved at ambient pressure after PQP, in comparison with the original $T_c$ ~ 133 K (right inset, Fig. 1B) for Hg1223 before PQP. An indication of possible superconductivity up to 172 K for S5, although not yet reproduced, was also included in Fig. S2 for completeness. Detailed results are discussed below.

Fig. 2A for Exp. 1 shows the R(T) for Hg1223 (S1) under different conditions: before PQP, curve 1—under 0.5 GPa, displaying a $T_c$ = 137 K on cooling and curve 2—under 26 GPa, showing a $T_c$ = 152 K on cooling, close to the $T_c$-P peak shown in Fig. 1B, targeted for PQP; and after PQP at $P_Q$ = 26 GPa and $T_Q$ = 77 K, curves 3 and 4 both on warming sequentially exhibiting a $T_c$ = 139 K, about 6 K higher than the ambient-pressure $T_c$ of 133 K for the pristine Hg1223 before loading into the DAC (right inset, Fig. 1B). Curve 3 was taken on warming from $T_Q$ = 77 K to 220 K immediately after PQP and curve 4 was taken on warming from 77 K after cooling from 220 K. For brevity and clarity, we have also summarized the above results, including the quench temperatures ($T_Q$s), quench pressures ($P_Q$s), targeted $T_c$s at $P_Q$s, and retained $T_c$s at ambient pressure after PQP, in Fig. 2F (Exp. 1).

We have shown (17) previously that by lowering the $T_Q$, the metastable superconducting phase with a higher $T_c$ may be retained more easily. We have, therefore, PQed Hg1223 (S1) at $P_Q$ = 28.4 GPa but at a lower $T_Q$ = 4.2 K. R(T) curves for Hg1223 (S1) under these conditions are displayed in Fig. 2B: before PQP, curve 5—exhibiting a $T_c$ = 154 K on cooling under 28.4 GPa; and after PQP, curves 6 and 7—displaying a $T_c$ = 147 K on warming from 4.2 K to ~ 165 K and on cooling from ~ 165 K, respectively. This shows that PQP at lower $T_Q$ has indeed retained the pressure-induced metastable superconducting phase with a higher ambient-pressure $T_c$ of 147 K, which is 14 K higher than the value of 133 K for the pristine sample at ambient pressure. The above results are summarized in Fig. 2F (Exp. 2).

To test the effect of $P_Q$ on the retained $T_c$ after PQP, we decided to lower it to ~ 10 GPa and at the same time reduce the possible pressure-shock effect, *e.g.* defects, on the transition associated with PQP, while maintaining the targeted $T_c$ close to or above 150 K, based on Fig. 1B, for PQP. The experiment shown in Fig. 2B was repeated but at a lower $P_Q$ = 10.1 GPa and at the same $T_Q$ = 4.2 K. Similar results were indeed achieved as shown in Fig. 2C: curve 8—at 10.1 GPa, showing a $T_c$ = 152 K on cooling targeted for PQP; and after PQP at 10.1 GPa and at $T_Q$ = 4.2 K, curves 9 and 10—warming from 4.2 K to ~ 160 K and cooling from ~ 160 K to 4.2 K, respectively, both displaying $T_c$ = 147 K, showing that, *via* PQP, the pressure-induced metastable superconducting phase has been retained with an ambient-pressure $T_c$ that is 14 K higher than the value of 133 K for the pristine sample at ambient pressure, and with a narrower superconducting transition consistent with that observed under pressure before PQP: $T_c$ = 152 K [$\Delta T$ = 9 K (Fig. 2C, curve 8)] *vs.* $T_c$ = 154 K [$\Delta T$ = 22 K (Fig. 2B, curve 5)], where $\Delta T = T_c - $ (T at R = 50% $R_{onset}$). Indeed, the retained superconducting transition width at ambient pressure is narrower than that shown in Fig. 2B: $\Delta T$ = 7 K (Fig. 2C, curve 9) *vs.* $\Delta T$ = 19 K (Fig. 2B, curve 6). The above results are summarized in Fig. 2F (Exp. 3). It appears that defects are reduced by reducing $P_Q$ (as reflected by a reduced $\Delta T$).

At a lower $P_Q$ = 10.1 GPa and $T_Q$ = 4.2 K, the PQP with a lower target $T_c$ ~ 152 K appears to retain the ambient-pressure $T_c$ at 147 K. We therefore increased the target $T_c$ to ~154 K for PQP by increasing $P_Q$ slightly to 10.9 GPa. R(T) curves for Hg1223 (S3) under different conditions are shown in Fig. 2D: prior to PQP, curve 11—under 10.9 GPa showing a $T_c$ = 154 K on cooling; and after PQP at $P_Q$ = 10.9 GPa and 4.2 K, curves 12 and 13—warming from 4.2 K to ~ 160 K and cooling from ~ 160 K to 4.2 K, respectively, showing respective $T_c$s of 149 K and 148 K. The above results are summarized in Fig. 2F (Exp. 4). We then further pushed the $P_Q$ to 18.9 GPa. R(T) curves for Hg1223 (S4) under different conditions are shown in Fig. 2E: prior to PQP, curve 14—under 18.9 GPa showing a $T_c$ = 158 K on cooling; and after PQP at $P_Q$ = 18.9 GPa and 4.2 K, curves 15 and 16—warming from 4.2 K to ~ 170 K and cooling from ~ 170 K to 4.2 K, respectively, showing respective $T_c$s of 151 K and 150 K. The above results are summarized in Fig. 2F (Exp. 5). They demonstrate that PQP has retained the pressure-induced metastable superconducting phase with a record-breaking ambient-pressure $T_c$ up to 151 K and suggest that higher $T_c$ will be attainable in HBCCO or similar compounds when the PQP technique is better refined.

It should be noted that although a new record-high $T_c$ of 151 K has been reproducibly achieved in Hg1223 at ambient pressure, the other retained $T_c$s are up to several K below the chosen target $T_c$ > 150 K for PQP. This may be attributed to the two different pressures to which the pre- and post-PQed samples were subjected, *i.e.* $P_Q$ and ambient pressure, respectively, and the positive $dT_c/dP$. The small spread of the retained $T_c$s (147–151 K) after PQP for $P_Q$s in the 10–30 GPa range and $T_Q$ = 4.2 K stems from the relatively flat $T_c$–P dependence above 10 GPa as shown in Fig. 1B.

Given our success in raising the ambient-pressure $T_c$, although only by 4 K, by increasing $P_Q$ to 18.9 GPa from 10.1 GPa, we therefore attempted to raise it further by increasing the $P_Q$ to 29.7 GPa. Sample S5 was PQed at $P_Q$ = 29.7 GPa (close to 30 GPa for maximum $T_c$, as shown in Fig. 1B) and $T_Q$ = 4.2 K, and the results are represented by three R(T) curves in Fig. S2: before PQP, 1—under 29.7 GPa, showing a $T_c$ = 156 K on cooling; and after PQP, 2—on warming from 4.2 K to 193 K, displaying a superconducting-like transition from ~ 110 K to 172 K, which also consists of two minor rapid rises at 138 K and 150 K, near the ambient-pressure $T_c$ before PQP and close to the $T_c$ at 29.7 GPa, respectively, and 3—on subsequent cooling from 180 K, the distinct drop below 172 K disappears although with a kink near 160 K followed by a drop below 136 K. It is possible that, following PQP, multiple phases with distinct $T_c$ values remain at ambient pressure, which is reasonable considering the “complex” shock nature of PQP and of the PQed micro-states. No amount of effort has been successful in reproducing the 172-K transition by PQP, and thus the true nature of the 172-K transition observed in Fig. S2, curve b, remains unresolved. However, in view of the possible thermal instability of the metastable phases at such a high temperature, the possibility that the PQed sample exhibits a high-temperature superconducting transition that then relaxes back to a lower $T_c$ phase cannot be completely ruled out, although more effort is needed to resolve this uncertainty. In fact, we have observed previously a higher-than-target $T_c$ in a PQed $Bi_{0.5}Se_{1.5}Te_3$ sample (28) and attributed this to a possible novel phase induced by the PQP shock effect.

Following our success in retaining at ambient pressure the high-pressure-induced high-$T_c$ phases, we decided to examine, within the confines of the DAC, some characteristics of the resistive transitions observed. As we have done previously in the search for high $T_c$, we needed to exclude possible artifacts, such as pressure-generated defects that could lead to the broadening of the transition and thus its apparent onset temperature rise, possibly not related to superconductivity. We have therefore determined the effect of applied current on the transitions we observed. As shown in Fig. S3, with increasing current, the entire resistive transition is shifted downward in temperature, characteristic of superconductivity. Therefore, the rapid rise of dR/dT *vs.* T, or $T_c$, observed by us and shown in Fig. 2, is indeed associated with the onset of a superconducting transition. The drop in R(T) with increasing current from 0.2 mA to 50 mA appears to be an artifact attributable to the possible local heating of the lead contacts and/or the relaxation of the sample’s normal-state resistance, with little effect on its $T_c$. This effect appears to be consistent with the lower resistance at 0.2 mA observed in later measurements.

**Stability Testing of Pressure-Quenched Hg1223**

To guarantee the successful characterization and application of the PQed phases, their complex thermal and temporal stability must be thoroughly examined. Preliminary temperature-dependent resistance data at ambient for the high-pressure-induced phases with a $T_c$ ~ 147 K in our PQed samples have been obtained during experimental cycles over different temperature spans and time periods. For instance, R(T) data obtained for Hg1223 (S2) following different thermal cycling conditions revealed a decrease in the retained ambient-pressure $T_c$ of 147 K to 143 K after we raised the maximum cycling temperature to room temperature Fig.S4). We have also demonstrated that the PQed phase with a $T_c$ = 149 K in Hg1223 (S3) is retained inside the DAC at 77 K for at least up to three days and is stable at temperatures up to 170 K (Fig. S5), although with a slight decrease in the retained ambient-pressure $T_c$. We tried to retrieve Hg1223 sample S2 from the DAC after PQ at 10.1 GPa and 4.2 K for DC magnetization measurements in a Quantum Design MPMS, as shown in Fig. S6. The results show that the superconductivity of the PQed phase is bulk in nature [~78%; details can be found in “Methods – Magnetization Measurements” (30)], although with a slightly lower $T_c$ of ~140 K, perhaps due to being warmed up to room temperature and/or partial thermal annealing of the sample during retrieval. The results show that the PQed superconductivity is not filamentary, as was also shown in our recent report on PQed $Bi_{0.5}Se_{1.5}Te_3$ (20), demonstrating the effectiveness of PQP for the retention of metastable phase materials for study and applications. As we have demonstrated, the PQed phase displays limited stability, both thermal and temporal. Further improvement of such stability to meet the requirements of both characterization by other probes (such as sample size, surface condition, *etc.*) and utilization of the metastable phases is crucial for the realization of the full potential of superconductivity and PQP.

**Crystal Structures and Possible Nanostructures of the PQed Hg1223**

Our synchrotron XRD results (Fig. 3A) show that Hg1223 crystalizes in a tetragonal structure (*P4mmm*) at ambient pressure and remains so to ~15 GPa (Fig. 3B), the highest pressure examined, in agreement with previous studies (31). This suggests that all the high $T_c$s of Hg1223 achieved under pressure were without any structural transition, which is consistent with the smooth $T_c(P)$ behavior observed. Given the fact that the $T_c$ of the PQed phase falls within the range defined by $T_c(P)$, the absence of a structural change induced by PQP, as indicated by the temperature-dependent XRD patterns shown in Fig. 3C, is not unexpected. However, a clear increase in the full width at half maximum (FWHM) of the (200) diffraction peak is detected in the PQed sample [0.20° *vs.* 0.12° for the pristine sample (Fig. 3D)]. A slightly enlarged lattice parameter was also detected in the PQed sample, indicating a negative pressure generated during PQP that could be related to the retention at ambient pressure of the high-pressure-induced metastable phase. To further test the hypothesis that strain and structural defects may play a key role, we plan to conduct systematic structural measurements on PQed samples exhibiting different superconducting $T_c$s under different controlled conditions. These studies will vary $T_Q$ and $P_Q$ parameters, thermal cycling protocols, and aging times to assess their impact on retained $T_c$ and phase stability. Investigation utilizing a hard X-ray nanoprobe under high-pressure conditions to study nanostructures in the sample may also be helpful. We believe that such nanostructures resulting from PQP may be responsible for the retention of the pressure-induced high-$T_c$ phase in Hg1223 at ambient pressure.

**Electron and phonon band structures of Hg1223**

To gain insight into the mechanism behind the rapid rise in $T_c$ for PQP with $P_Q$ ~10 GPa, we computed the band structure of Hg1223 under pressure. We did not see any structural transition up to ~50 GPa, indicated by the absence of negative phonon modes (Fig. S7), consistent with our experiments. However, we saw an electronic transition around ~12.5 GPa, as shown in Fig. 4. Specifically, we see a Lifshitz transition—the appearance of new Fermi pockets composed of 2p electrons from the apical oxygen (Fig. S8)— that leads to a sudden rise in the density of states, giving rise to a high $T_c$ under pressure. Regardless of the pairing symmetry, correlations, and other details of superconductivity, a large rise in the normal density of states is generally expected to cause a sharp increase in $T_c$. Thus, we speculate that states with and without these Fermi pockets may be separated by an energy barrier that renders the higher-pressure phase metastable at ambient pressure, enabling the retention at ambient pressure of the record high-$T_c$ superconductivity (excluding that of the hydrides, which are unstable at ambient pressure).

**Conclusion**

Here we report the reproducible attainment of a record-high $T_c$ up to 151 K in bulk tetragonal Hg1223 at ambient pressure by employing our novel pressure-quench protocol (PQP) at $P_Q$ ~ 10.1 – 28.4 GPa and $T_Q$ = 4.2 K, representing an increase up to 18 K above the previous record of 133 K (8). Synchrotron XRD data show that the PQed phase at ambient pressure remains in its original crystal structure but displays a large line-broadening, attributable to the presence of defects generated under pressure and during PQ. We believe that these defects help retain the metastable high-$T_c$ phase, evidenced by the sharpening of the superconducting transition and the lowering of the $T_c$ through annealing to a higher temperature or for a longer period. Additionally, oxygen content and/or vacancies in Hg1223 (32) are known to strongly influence its superconducting properties. We plan to perform careful microscopy studies (33) on both pristine and pressure-quenched samples to identify the potential role of oxygen content and oxygen vacancies in stabilizing the pressure-enhanced $T_c$ through the PQP. Evidence shows that, after perfecting the PQP process, it may be possible to achieve a $T_c$ in Hg1223 above 151 K, in view of our anomalous observation at ~172 K. We also found that the retained ambient-pressure $T_c$ decreases to 139 K when $T_Q$ is increased to 77 K, only ~6 K above that prior to PQP at ambient pressure, in agreement with our previous studies. The superconducting phase achieved *via* PQP is bulk (~78%) in nature, as evident from our DC magnetization measurements on a sample retrieved from the DAC after PQP. The metastable superconducting phase retained at ambient pressure is found to be stable for at least three days when kept in liquid nitrogen at 77 K, but its $T_c$ degrades when heated to above 200 K. The success of this experiment demonstrates the great potential of the PQP technique in stabilizing high-pressure-induced metastable states with enhanced properties at ambient pressure for scientific study and practical application, even beyond the current discovery of superconductivity with record-high $T_c$. Further detailed studies focusing on the specific parameters that optimize the PQP process and the characteristics to define the PQed product are necessary to fully realize the immense potential of PQP for the generation, characterization, and utilization of the unique materials with desired characteristics so-produced under pressure. The parameters to be studied and refined include quenching pressure ($P_Q$), quenching temperature ($T_Q$), quenching rate ($v_Q$), and the thermodynamics and kinematics of PQP, while the product characteristics consist of all those that can be measured either inside or outside the DAC using currently available techniques, especially the micro- and nanostructures to be determined by synchrotron XRD, Raman spectroscopy, the nitrogen-vacancy (NV) center technique, differential scanning calorimetry (DSC), and differential thermal analysis (DTA). A direct comparison between the PQed sample extracted from the DAC and the pristine sample using STM (23) and angle-resolved photoemission spectroscopy (ARPES) will provide critical insight into the microscopic mechanism underlying the high $T_c$ in this system, as well as the success of PQP. We thus believe that the record-breaking results at ambient pressure reported here represent only the beginning of an extremely fruitful scientific journey.

## Materials and Methods

**Sample Preparation.** Single crystals of $HgBa_2Ca_2Cu_3O_{8+\delta}$ were synthesized using the self-flux method (34). Powders of HgO (99.98%, Alfa Aesar), BaO (99.95%, Alfa Aesar), CaO (99.95%, Alfa Aesar), and CuO (99.99%, Alfa Aldrich) were utilized as starting materials. To ensure material integrity and prevent contamination, all handling of the oxides was conducted in an argon-filled glove box with $O_2$ < 0.1 ppm and $H_2O$ < 0.1 ppm. The mixture of raw materials was pressed into pellets and then heated at a rate of 200 °C/h to 725 °C, which was maintained for 1 h, followed by rapid heating (300 °C/h) to 860 °C and soaking for 3 h. The temperature was then lowered to 600 °C at a rate of 1 °C/h, and the furnace was subsequently cooled down with a setting of 200 °C/h to room temperature. Post annealing under oxygen flow was carried out at 325 °C for 5–10 days to achieve optimal ambient-pressure $T_c$ (above 130 K). X-ray diffraction measurements were performed *via* a Rigaku X-ray diffractometer equipped with a $CuK\alpha_1$ X-ray source (λ = 1.54059 Å) at room temperature to ensure the correct phase.

**Transport Measurements under Pressure.** For resistivity measurements conducted in this investigation, pressure was applied to the samples using Mao-type symmetric DACs (35) with culet sizes of 300 μm and 400 μm. The gaskets are made from T301 half-hard stainless-steel sheets with thickness of 300 μm. Each gasket was preindented to ~ 20 – 40 μm in thickness and was insulated with Stycast 2850FT. The sample's chamber diameter is ~100 – 130 μm, where cubic boron nitride is used as the pressure-transmitting medium. Samples were cleaved and cut into thin squares with a diagonal of ~ 80 – 120 μm and thickness of ~ 20 μm. The pressure was determined using the ruby fluorescence scale (36) or the diamond Raman scale (37) at room temperature. The samples' contacts were arranged in a Van der Pauw configuration, and data were collected using a Keithley 6221/2182A Delta Mode System. Measurements were conducted in a homemade cooling system that can be cooled to 1.2 K by pumping on the liquid-helium space. PQP was performed by releasing the screws at target temperatures down to 4.2 K.
**Magnetization Measurements**. Magnetization measurements from 70 K to 160 K under 10 Oe were carried out to determine the $T_c$ of Hg1223 samples before and after PQP using a Quantum Design Magnetic Property Measurement System (MPMS 3). Superconducting volume fraction for a thin-disk-shaped sample can be calculated (30) using $f = 4\pi\chi = \frac{4\pi|M|}{HVD} \approx \frac{4\pi|M|}{H\times(\pi r^2 d)\times(\frac{4r}{\pi d})} = \frac{\pi|M|}{Hr^3}$, where *M* is the magnetic moment, *H* is the applied magnetic field, *V* is the sample volume, *D* is the demagnetization factor, *r* is the radius of the cross section of the sample, and *d* is the sample thickness. In our case, $|M|/H = 1.9\times10^{-8}$ emu/Oe at 70 K and $r \approx 42.5\times10^{-4}$ cm. *f* is estimated to be 78%.
**High-Pressure X-ray Diffraction Measurements.** High-pressure single-crystal X-ray diffraction (XRD) measurements were conducted at the 16-ID-B beamline of the Advanced Photon Source (APS), operated by the High-Pressure Collaborative Access Team (HPCAT). The experimental setup utilized a monochromatic X-ray beam ($\Delta E/E$, $1\times10^{-4}$) with the energy of 29.200 keV ($\lambda$ = 0.4246 Å), focused to spot sizes as small as 2×2 $\mu m^2$ using Kirkpatrick–Baez (KB) mirrors. This configuration provides high photon flux, approximately 5 $\times10^{12}$ photons per second at 29.200 keV, ensuring sufficient intensity for detecting weak diffraction signals from small single crystals. Hg1223 single-crystal samples were loaded into DACs that are compatible with the beamline's goniometer systems, allowing precise orientation and rotation of the crystal under pressure. Pressure within the DACs was monitored using online ruby fluorescence systems, facilitating accurate determination of the pressure conditions during measurements. High-pressure XRD patterns were collected using a PILATUS3 X 2M CdTe area detector. The experimental station supports angle-dispersive XRD techniques, enabling detailed structural analysis of single crystals under varying pressure conditions. The beamline equipped with a DAC-compatible cryostat was used for temperature control over the range of 20–300 K.
**Density Functional Theory.** Density functional theory calculations were carried out using the projector augmented wave (PAW) method within the Vienna Ab initio Simulation Package (VASP) (38–40). The exchange-correlation interactions were treated using the Perdew–Burke–Ernzerhof (PBE) form of the generalized gradient approximation (GGA) (41). Hydrostatic pressure was applied up to 50 GPa, and at each pressure point, both the atomic positions and lattice parameters were fully relaxed. The electronic band structures were further computed using the full-potential local orbital (FPLO) code, version 22.00 (42), employing energy and charge density convergence criteria of $10^{-8}$ Hatree and $10^{-6}$ e/(Bohr radii) (40), respectively. A Γ-centered 24 × 24 ×12 *k*-point mesh was used to sample the Brillouin zone for self-consistent and band structure calculations, whereas for the Fermi surface, a mesh of 40 × 40 × 40 was used. To study the lattice dynamics, phonon dispersion relations were calculated using the finite-displacement method implemented in the PHONOPY package (43). A 2×2×1 supercell was used in these calculations to obtain the required force constants. The initial lattice constants in our calculations and effects of pressure on them are well matched with previous reports (31,44).

**Acknowledgments**

L. Z. D., T. H., A. S., S. K., D. J. S., Z. W., and C. W. C. in Houston are supported in part by the Enterprise Science Fund of Intellectual Ventures Management, LLC; U.S. Air Force Office of Scientific Research Grants FA9550-15-1-0236 and FA9550-20-1-0068; the T. L. L. Temple Foundation; the John J and Rebecca Moores Endowment; and the State of Texas through the Texas Center for Superconductivity at the University of Houston. L. Z. D. in Houston is supported in part by the Robert A. Welch Foundation (00730-5021-H0452-B0001-G0512489). M. J. and R. P. P. are supported by the Enterprise Science Fund of Intellectual Ventures Management, LLC. B. K. and P. H. are supported by the State of Texas through the Texas Center for Superconductivity. This research was performed on APS beam time award(s) (DOI:10.46936/APS-188998/60013464) from the Advanced Photon Source, a U.S. Department of Energy (DOE) Office of Science user facility operated for the DOE Office of Science by Argonne

National Laboratory under Contract No. DE-AC02-06CH11357. HPCAT operations are supported by DOE-NNSA's Office of Experimental Sciences. We also acknowledge helpful discussions with Dr. Nathan Myhrvold.

## Figures

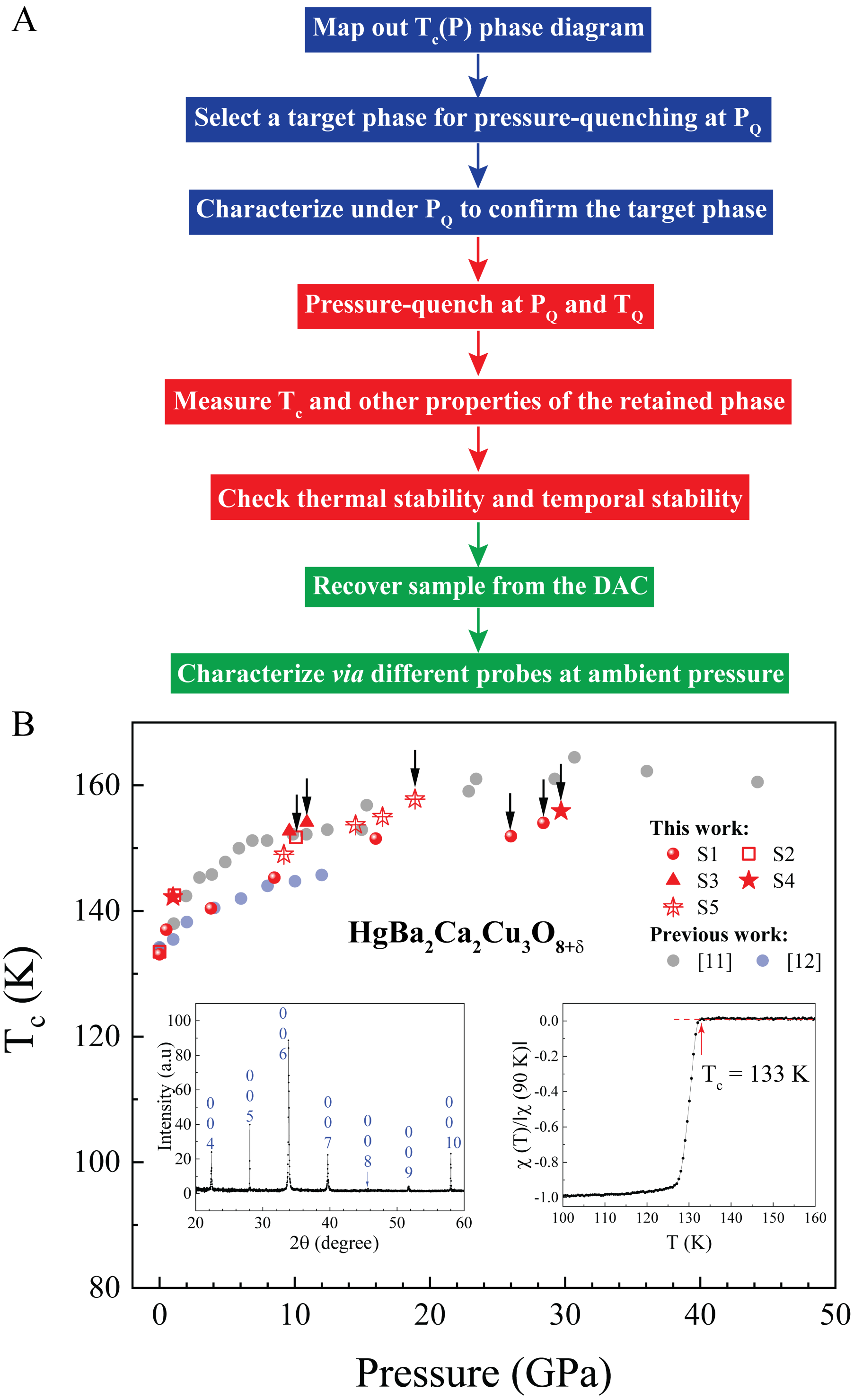

**Figure 1.** (A) Schematic of the pressure-quench protocol (PQP). $P_Q$, $T_Q$, and DAC: quenching pressure, quenching temperature, and diamond anvil cell, respectively. Different colors indicate the three primary stages of PQP as discussed in the main text. (B) Pressure dependence of the onset $T_c$ for Hg1223. Red symbols: this work. Gray and blue circles: Refs. 11 and 12, respectively. The arrows indicate the target states for PQP discussed in this work. Insets: XRD (left) and χ(T) (right) results for pristine Hg1223 (crystal #1) at ambient pressure.

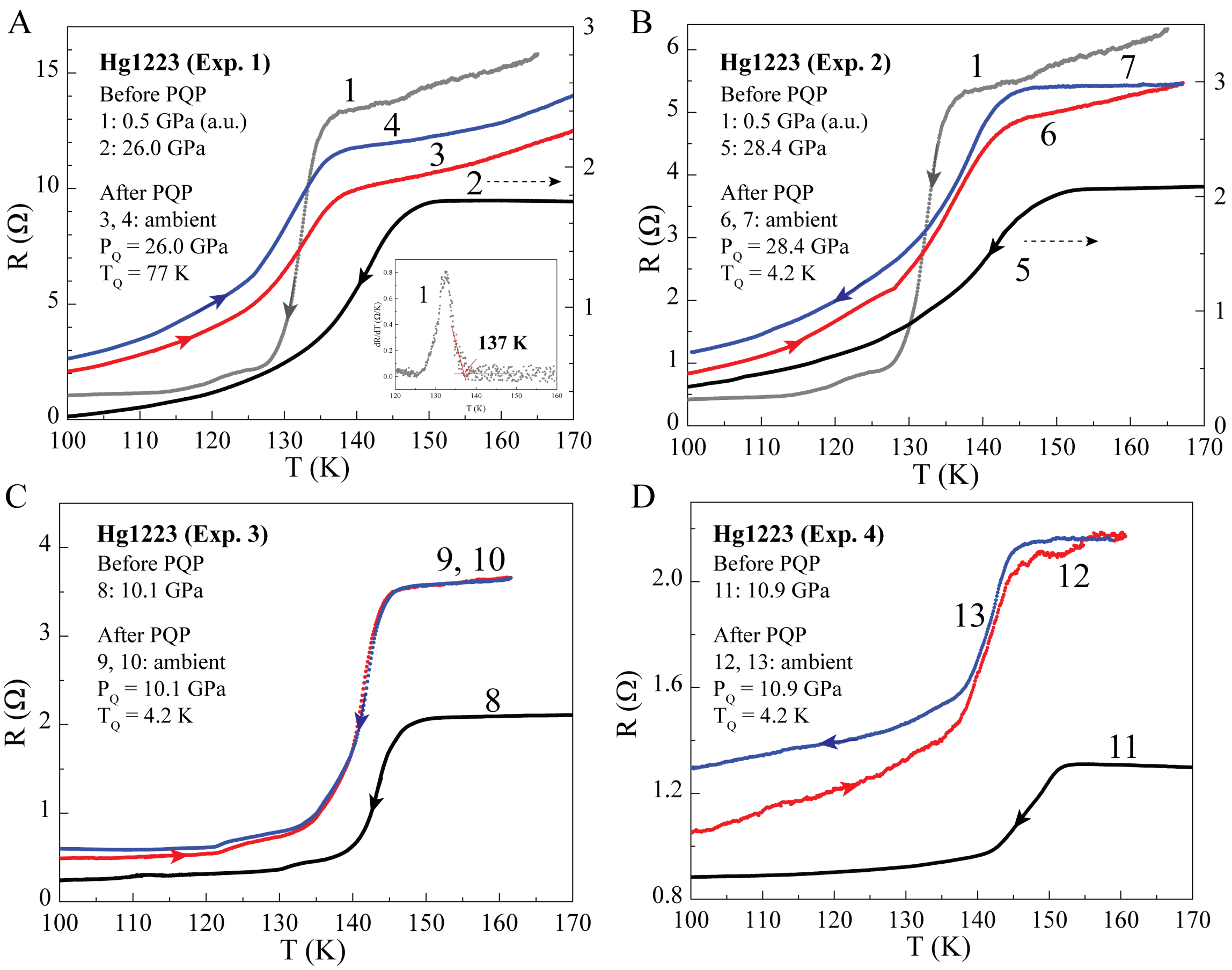

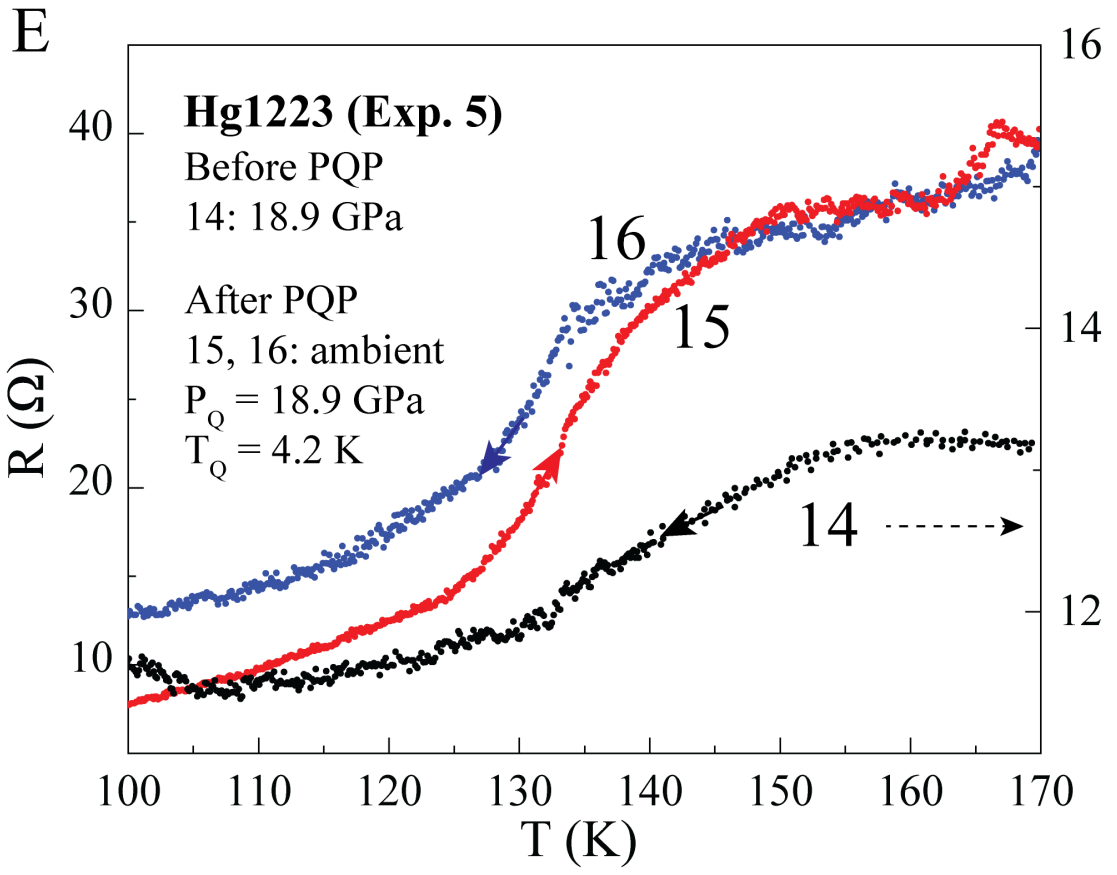

F

| Exp. | 1 | 2 | 3 | 4 | 5 |
|---|---|---|---|---|---|
| Sample | S1 | S1 | S2 | S3 | S4 |
| $T_Q$ (K) | 77 | 4.2 | 4.2 | 4.2 | 4.2 |
| $P_Q$ (GPa) | 26.0 | 28.4 | 10.1 | 10.9 | 18.9 |
| Target $T_c$ (K) at $P_Q$ | 152 | 154 | 152 | 154 | 158 |
| Retained $T_c$ (K) at ambient P | 139 | 147 | 147 | 149 | **151** |

**Figure 2.** Temperature-dependent resistance R(T) for Hg1223 under different conditions before and after PQP. (A) Before PQP, 1—under 0.5 GPa on cooling and 2—under 26.0 GPa on cooling; and after PQP at $P_Q$ = 26.0 GPa and $T_Q$ = 77 K, 3 and 4—on warming. (B) Before PQP, 5—under 28.4 GPa on cooling; and after PQP at $P_Q$ = 28.4 GPa and $T_Q$ = 4.2 K, 6—on warming and 7—on cooling. (C) Before PQP, 8—under 10.1 GPa on cooling; and after PQP at $P_Q$ = 10.1 GPa and $T_Q$ = 4.2 K, 9—on warming and 10— on cooling. (D) Before PQP, 11—under 10.9 GPa on cooling; and after PQP at $P_Q$ = 10.9 GPa and $T_Q$ = 4.2 K, 12—on warming and 13—on cooling. (E) Before PQP, 14—under 18.9 GPa on cooling; and after PQP at $P_Q$ = 18.9 GPa and $T_Q$ = 4.2 K, 15—on warming and 16—on cooling. Curves 2, 5, and 14 use the right y-axis scale as indicated by the arrows. The inset to (A) shows the temperature-dependent dR/dT results for curve 1 and the method for determining the corresponding onset $T_c$, while those for curves 2–16 are shown in Fig. S2. (F) Summary of the main parameters for PQP results for Exp. 1–5.

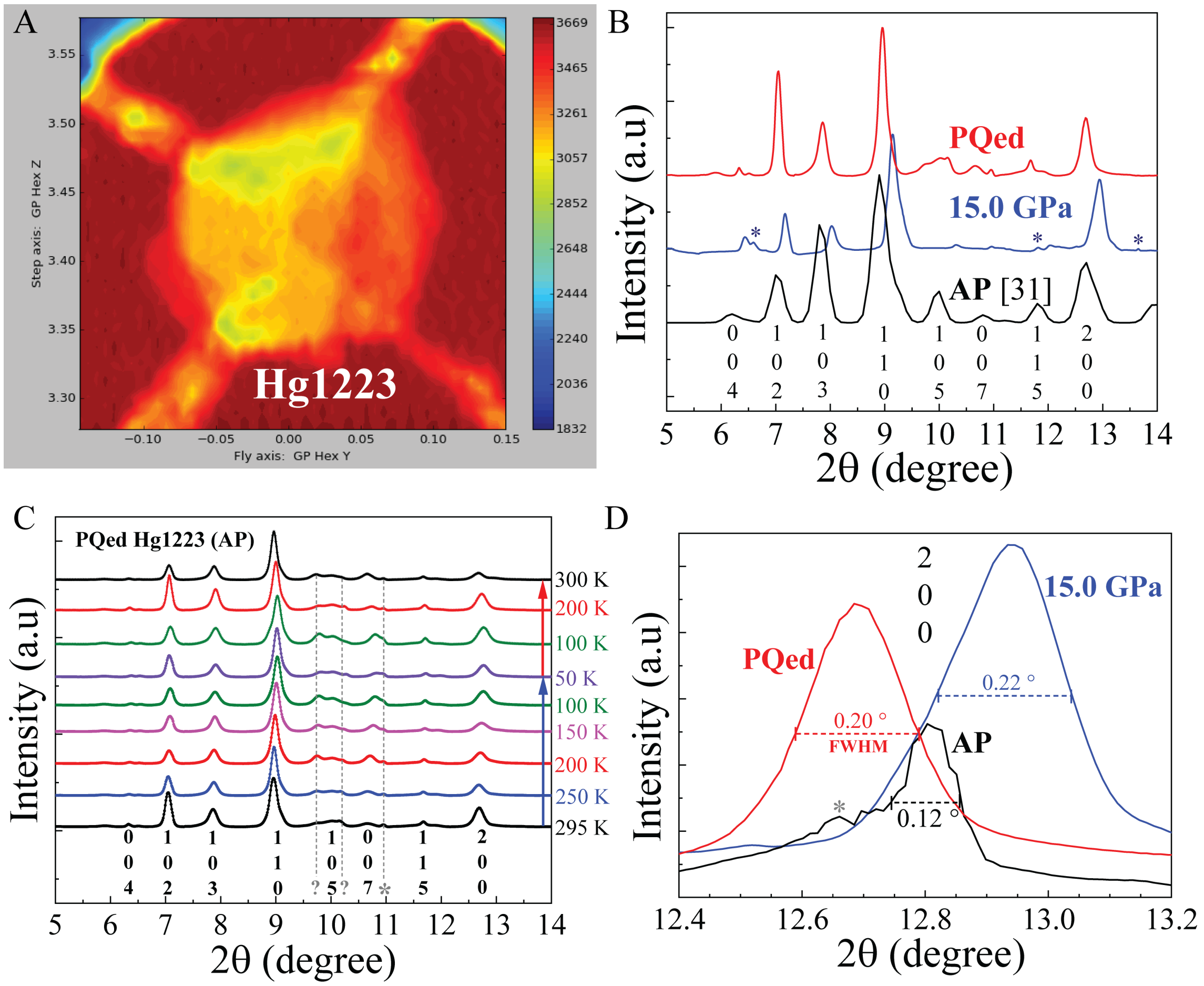

**Figure 3.** Synchrotron X-ray analysis of Hg1223 samples. (A) X-ray illumination absorption two-dimensional (2D) mapping of a Hg1223 sample inside a DAC with four Pt leads attached to the sample. (B) Pressure-dependent XRD patterns for Hg1223. (C) Temperature-dependent XRD patterns for PQed Hg1223 at ambient pressure. (D) (200) peak of Hg1223 at ambient pressure (black), under 15 GPa (blue), and at ambient pressure after PQ (red). Peaks labeled with "*" show temperature independence and are not from the sample but rather from the cryostat, *etc.* Peaks labeled with "?" are non-identified peaks. AP: ambient pressure. FWHM: full width at half maximum.

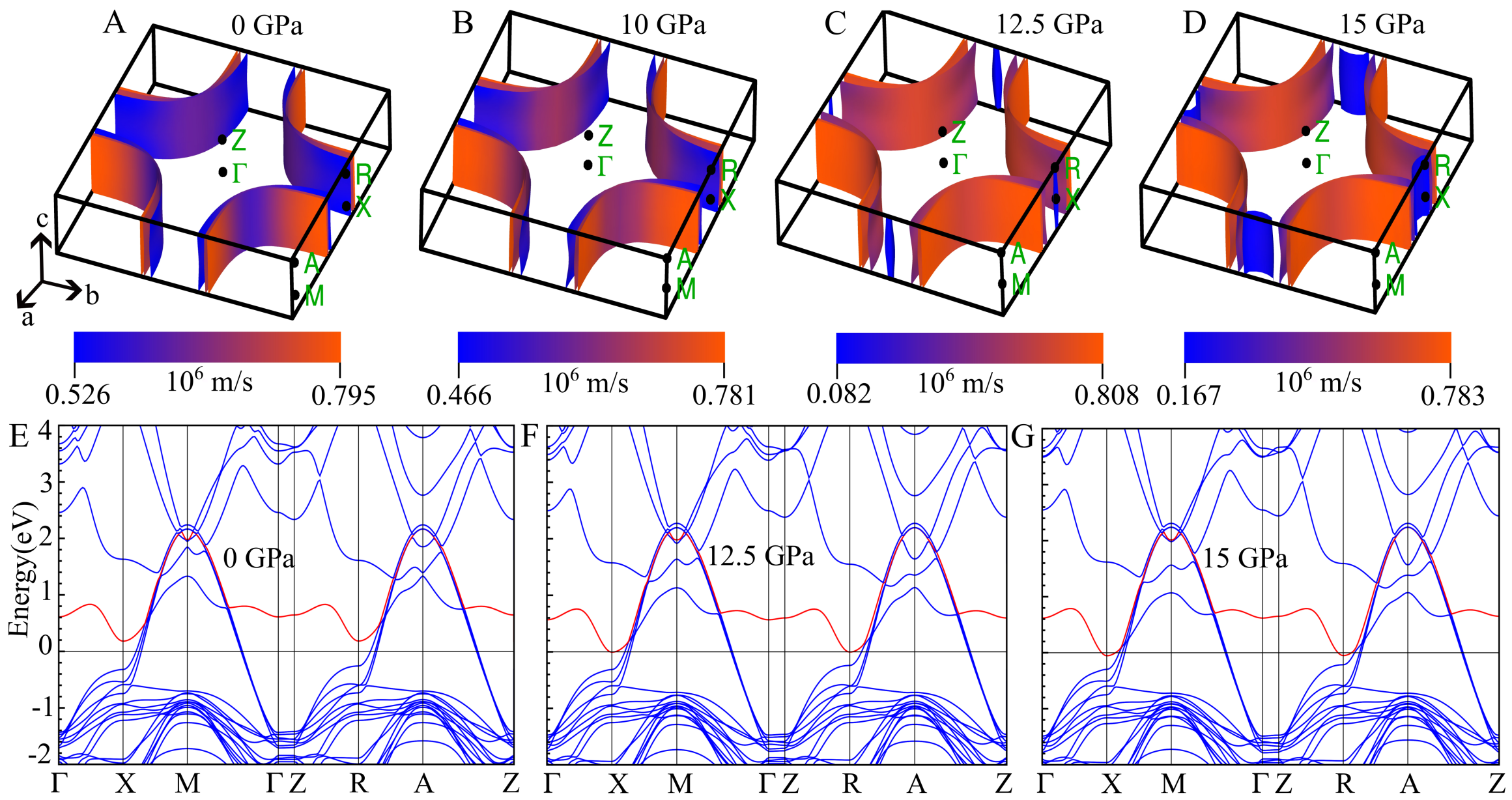

**Figure 4.** Electronic structures of Hg1223 from 0 GPa to 15 GPa. The Fermi level is aligned to zero energy in the band structure plots, which are represented with solid lines. (A-D) Evolution of the Fermi surface from 0 to 15 GPa. The color scale represents the magnitude of the Fermi velocity. A clear modification in the Fermi surface topology is observed at pressure from 12.5 GPa. (E) Electronic band structure of Hg1223 at ambient pressure. The red-colored band denotes the lowest conduction band in the electronic region, which lies entirely above the Fermi level and does not contribute to it at this pressure. This band is tracked to observe pressure-induced changes at the Fermi level. (F) Electronic band structure at 12.5 GPa. The same red band now touches the Fermi level, indicating its contribution to the electronic states at the Fermi energy. (G) Electronic band structure at 15 GPa, with the red band forming the Fermi pocket.

**Supporting Information for**

# Ambient-pressure 151-K superconductivity in $HgBa_2Ca_2Cu_3O_{8+\delta}$ *via* pressure quench

Liangzi Deng[1,*], Thacien Habamahoro[1], Artin Safezoddeh[1], Bishnu Karki[1], Sudaice Kazibwe[1], Daniel J. Schulze[1], Zheng Wu[1], Matthew Julian[2], Rohit P. Prasankumar[2], Hua Zhou[3], Jesse S. Smith[3], Pavan R. Hosur[1], Ching-Wu Chu[1,*]

Liangzi Deng, Ching-Wu Chu
Email: cwchu@uh.edu, ldeng2@central.uh.edu

## Figures

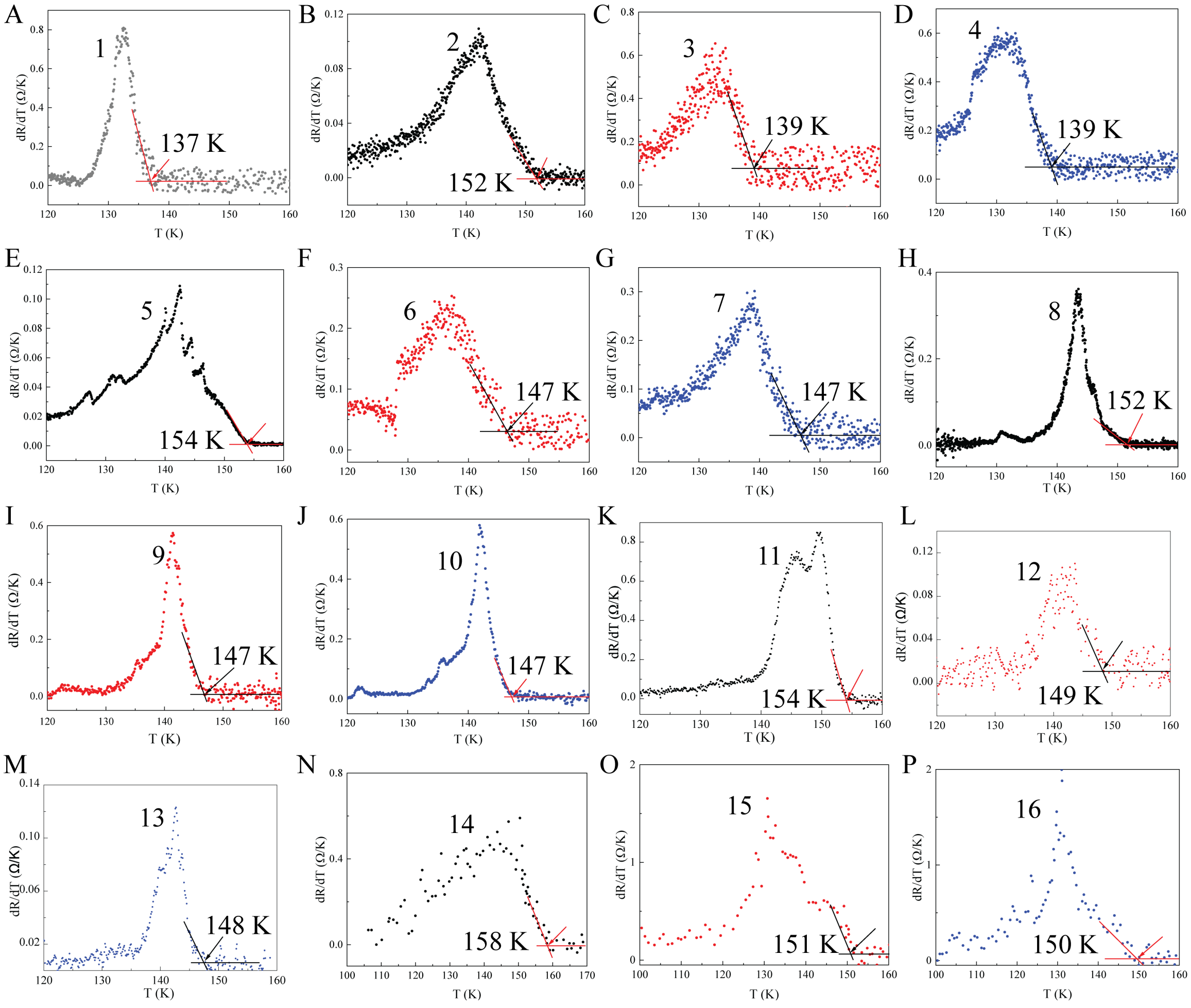

**Fig. S1.** (A–P) Temperature-dependent dR/dT results for R(T) curves 1–16, respectively, in Fig. 2, which are used for determining the corresponding onset $T_c$s.

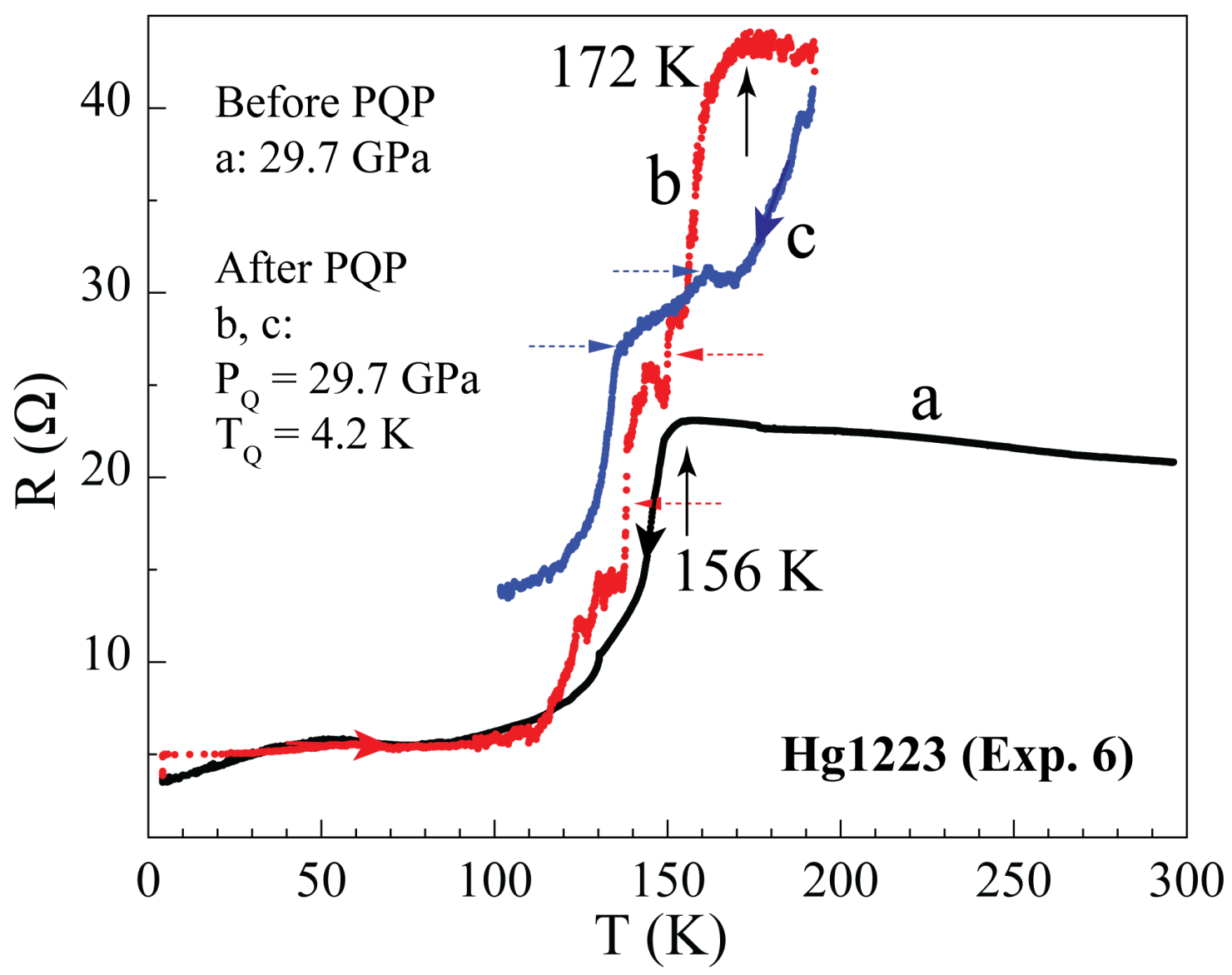

**Fig. S2.** Temperature-dependent resistance R(T) for Hg1223 (S5) under different conditions before and after PQP. a—before PQP under 29.7 GPa on cooling; and after PQP at $P_Q$ = 29.7 GPa and $T_Q$ = 4.2 K, b—on warming and c—on cooling. Dashed arrows indicate possible superconducting transitions retained after PQP during different thermal cycles.

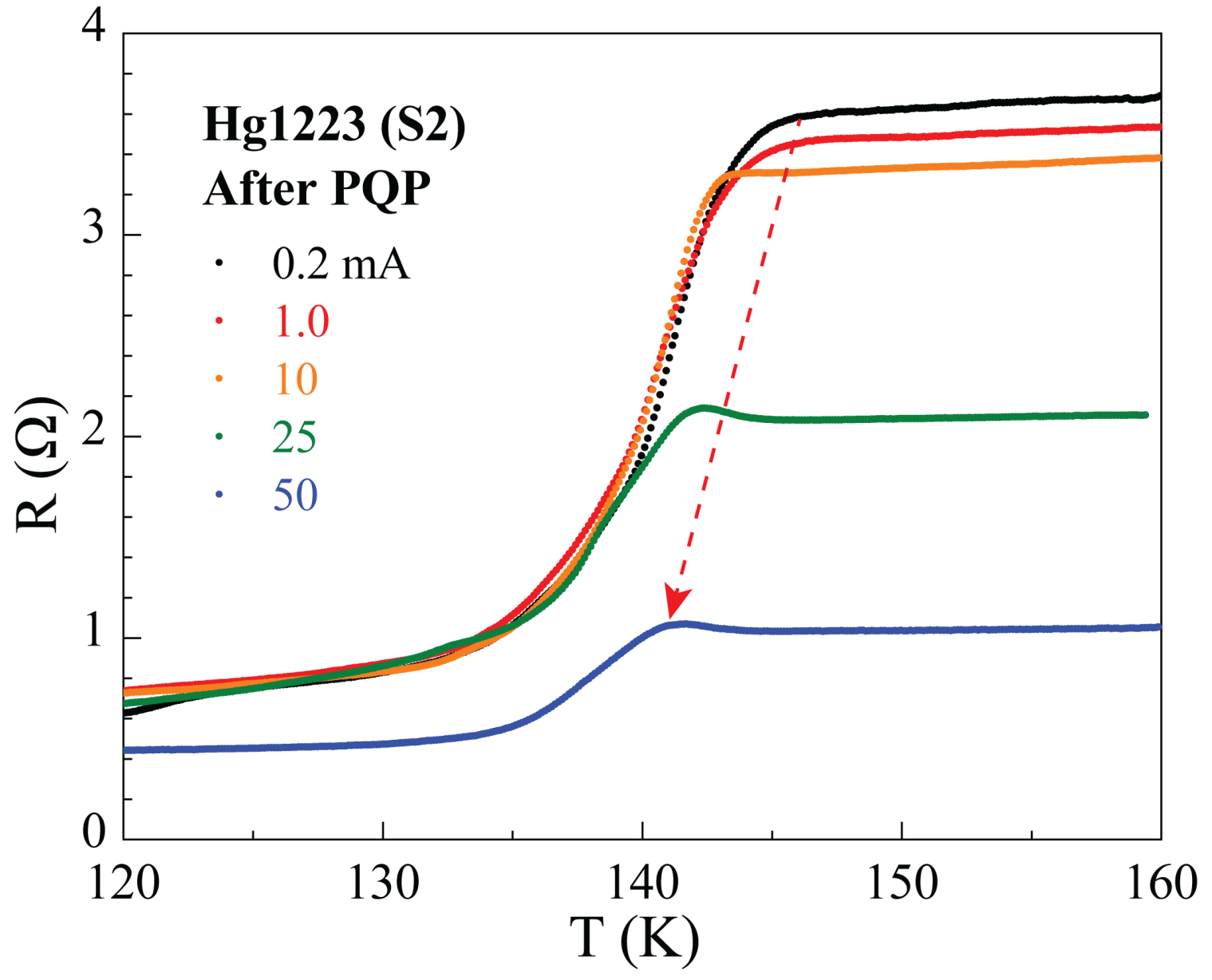

**Fig. S3.** Parallel downward shifting of R(T) with increasing current confirms the superconducting nature of the transition for Hg1223 sample S2.

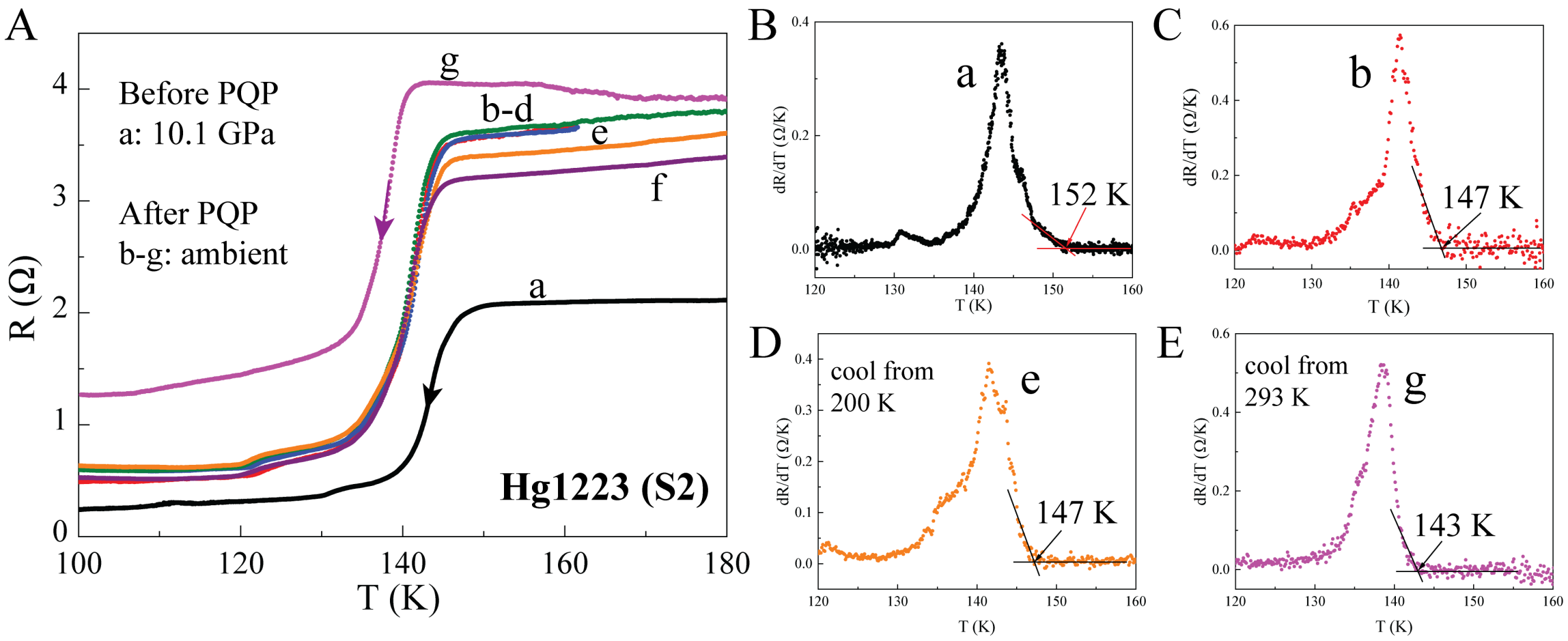

**Fig. S4.** Thermal stability testing of Hg1223 (S2). (A) a—before PQP under 10.1 GPa on cooling; and after PQP at $P_Q$ = 10.1 GPa and $T_Q$ = 4.2 K, b—on warming, c—on cooling from 160 K, d—on warming, e—on cooling from 200 K, f—on warming, and g—on cooling from 293 K (R/3). Panels at right (B–E): corresponding dR/dT *vs.* T data to determine the $T_c$s for curves a, b, e, and g, respectively.

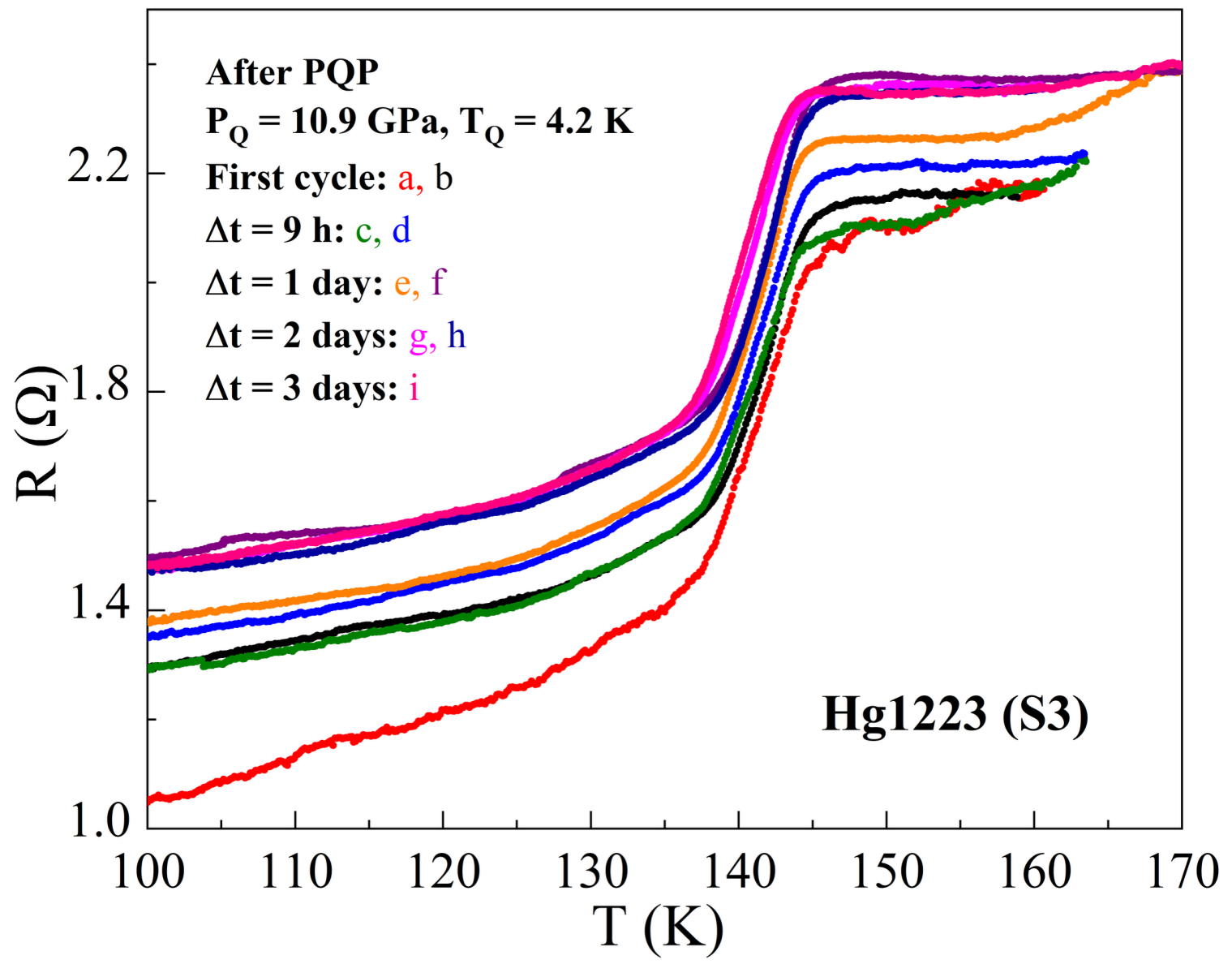

**Fig. S5.** Temporal stability testing of Hg1223 (S3). Immediately after PQP at $P_Q$ = 10.9 GPa and $T_Q$ = 4.2 K, a—on warming to 160 K and b—on cooling; 9 hours after PQP, c—on warming to 160 K and d—on cooling; 1 day after PQP, e—on warming to 170 K and f—on cooling; 2 days after PQP, g—on warming to 160 K and h—on cooling; 3 days after PQP, i—on warming to 170 K.

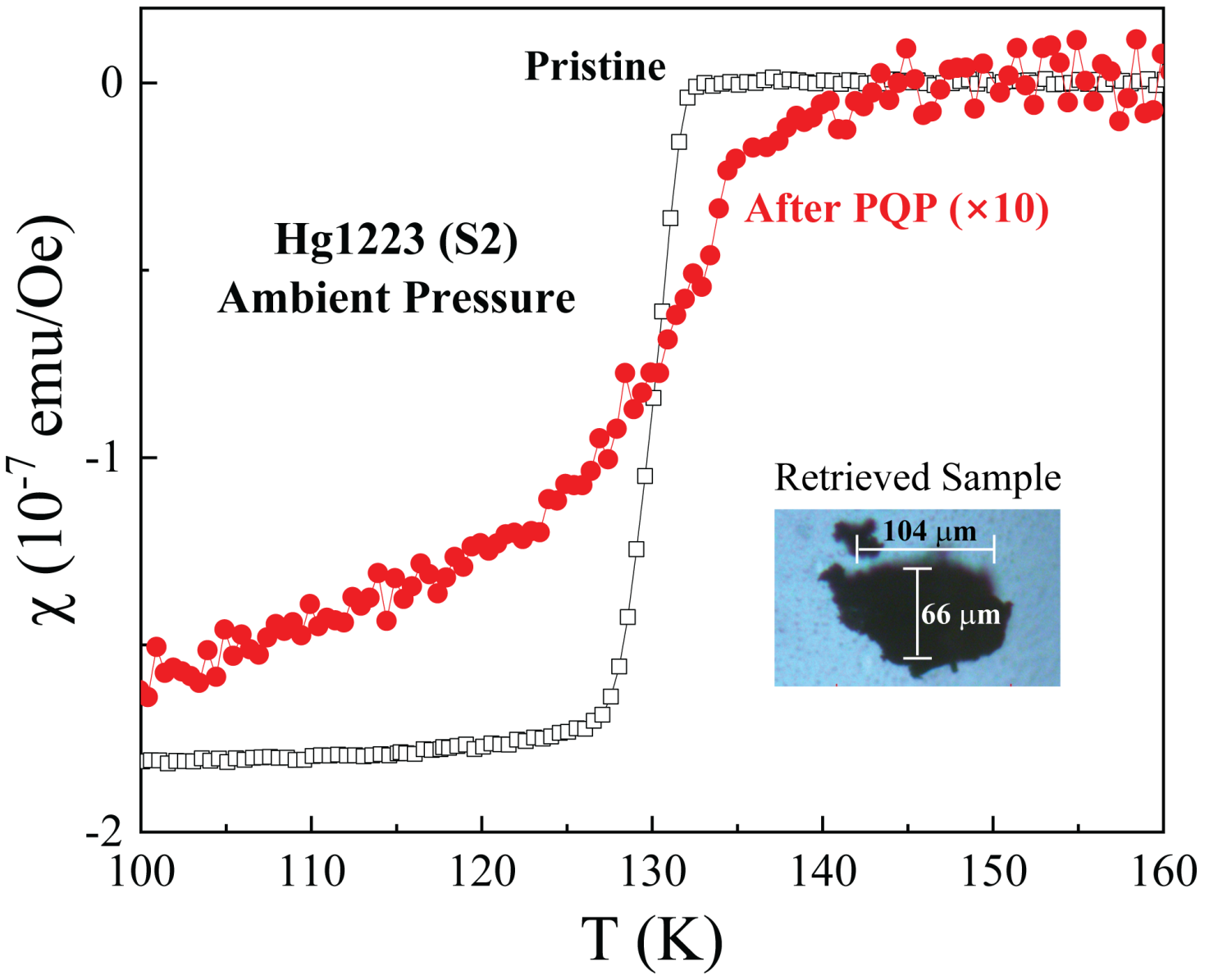

**Fig. S6.** DC magnetic results for pristine Hg1223 crystal #1 (source of S2) before PQP (black squares) and for Hg1223 sample S2 following retrieval from the DAC after PQP (red circles). Inset: part of Hg1223 (S2) retrieved from the DAC after PQP.

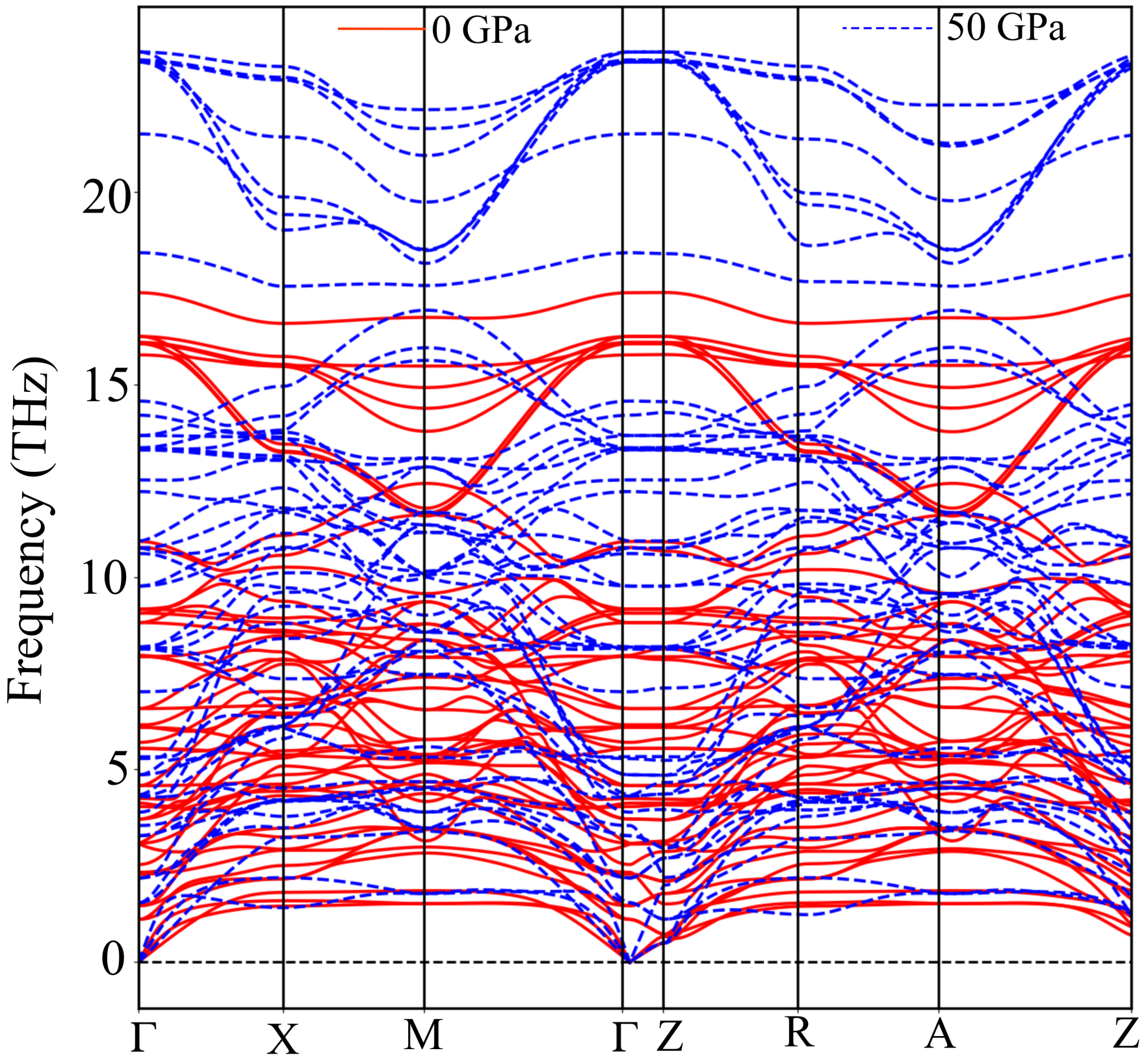

**Fig. S7.** Comparison of phonon dispersion in Hg1223 between 0 and 50 GPa. The absence of imaginary modes indicates structural stability without any structural transition up to 50 GPa.

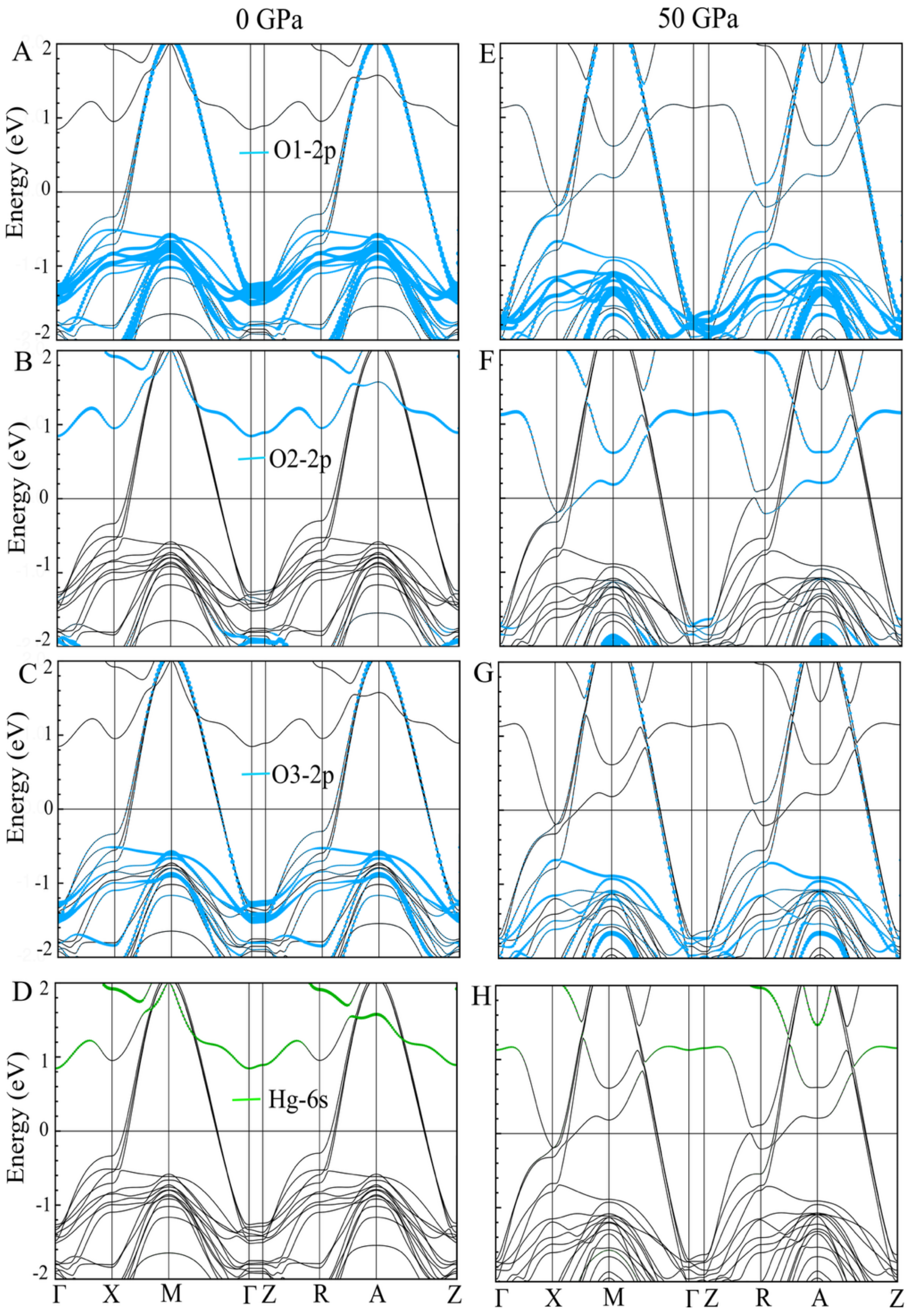

**Fig. S8.** Projection of orbital contributions on DFT bands from different oxygen sites (O1 and O3 from the Cu-O planes and O2 being the apical oxygen) and from Hg under ambient pressure and under high pressure. (A and E) O1, (B and F) O2, (C and G) O3, and (D and H) Hg for 0 GPa and 50 GPa, respectively. By comparing (B) and (F), it is clear that the O2-2p orbitals contribute at the Fermi level under high pressure.